\documentclass[12pt]{article}
\usepackage{physics}
\usepackage{epsf}
\usepackage{aas_macros}
\usepackage{amssymb}
\usepackage{comment}
\usepackage{amsmath}
\usepackage{braket}
\usepackage{mathrsfs}
\usepackage{mathtools}
\usepackage{array}
\usepackage{fancyhdr}
\usepackage[dvipdfmx]{graphicx}
\usepackage{color}
\usepackage{cite}
\usepackage{bm}
\usepackage{url}
\usepackage{fancyhdr}
\usepackage[colorlinks,citecolor=blue]{hyperref}
\usepackage[hang,small,bf]{caption}
\usepackage[subrefformat=parens]{subcaption}
\renewcommand{\thefootnote}{\#\arabic{footnote}}

\renewcommand{\thefootnote}{\fnsymbol{footnote}}
\def\thefootnote{\fnsymbol{footnote}}

\makeatletter

\@addtoreset{equation}{section}
\makeatother

\def\be{\begin{equation}}
\def\ee{\end{equation}}
\def\ben{\begin{eqnarray}}
\def\een{\end{eqnarray}}

\begin{document}


\begin{center}

\vskip .75in

{\Large \bf Upper limit on scalar field dark matter from LIGO-Virgo third observation run}

\vskip .75in

{\large
Koki Fukusumi$\,^1$,  Soichiro Morisaki$\,^2$,  Teruaki Suyama$\,^1$
}

\vskip 0.25in

{\em
$^{1}$Department of Physics, Tokyo Institute of Technology, 2-12-1 Ookayama, Meguro-ku,
Tokyo 152-8551, Japan \\
$^{2}$Institute for Cosmic Ray Research, The University of Tokyo,
5-1-5 Kashiwanoha, Kashiwa, Chiba 277-8582, Japan
}

\end{center}
\vskip .5in

\begin{abstract}
If dark matter is a light scalar field weakly interacting with elementary particles,
such a field induces oscillations of the physical constants, which results in
time-varying force acting on macroscopic objects.
In this paper, we report on a search for such a signal in the data of the two LIGO detectors 
during their third observing run (O3). 
We focus on the mass of the scalar field in the range of $10^{-13}-10^{-11}~{\rm eV}$ 
for which the signal falls within the detectors’ sensitivity band.
We first formulate the cross-correlation statistics that can be readily compared 
with publically available data.
It is found that inclusion of the anisotropies of the velocity distribution of dark matter 
caused by the motion of the solar system in the Milky Way Galaxy
enhances the signal by a factor of $\sim 2$ except for the narrow mass range around
$\simeq 3\times 10^{-13}~{\rm eV}$ for which the correlation between the interferometer
at Livingston and the one at Hanford is suppressed.
From the non-detection of the signal, we derive the upper limits on the coupling constants
between the elementary particles and the scalar field for five representative cases.
For all the cases where the weak equivalence principle is not satisfied,
tests of the violation of the weak equivalence principle provide the tightest upper limit
on the coupling constants.
Upper limits from the fifth-force experiment are always stronger than the ones
from LIGO, but the difference is less than a factor of $\sim 5$ at large-mass range.
Our study demonstrates that gravitational-wave experiments are starting to bring
us meaningful information about the nature of dark matter.
The formulation provided in this paper may be applied to the data of upcoming 
experiments as well and is expected to probe much wider parameter range of the model.
\end{abstract}

\renewcommand{\thepage}{\arabic{page}}
\setcounter{page}{1}
\renewcommand{\thefootnote}{\#\arabic{footnote}}
\setcounter{footnote}{0}

\section{Introduction}
Although there is little doubt for the existence of dark matter,
it's nature is still elusive.
There is considerable uncertainty in the mass of constituting elementary particle 
or object:
the allowed mass range is $10^{-19}~{\rm eV} \lesssim m \lesssim 10^{60}~{\rm eV}$.
Many models of dark matter such as Weakly Interacting Massive Particles (WIMPs), 
axions, and primordial black holes (PBHs) have been proposed in the literature.
Different models predict different observational signals and various experiments
suited for detecting specific type of dark matter have been conducted.
Yet, no robust detection is reported (for a review of dark matter, see e.g. \cite{Arbey:2021gdg}).
In light of this situation, it is important to consider utilizing 
even available experiments whose main target is not dark matter.

A dark matter model we consider in this paper is a light scalar field 
dark matter weakly interacting with elementary particles in the standard 
model of particle physics.
Such a scalar field may be realized in string landscape (see e.g., \cite{Kaplan:2000hh, Arvanitaki:2014faa}), although
seeking the fundamental origin is not our focus and we 
remain at the phenomenological level in this paper.
If the mass of the boson is $m\ll 10~{\rm eV}$, the dark matter behaves
as a classical wave \cite{Hui:2021tkt}.
Such a wave induces time variation of the physical constants and 
leads to detectable signals such as in the atomic transitions \cite{Derevianko:2013oaa, Arvanitaki:2014faa, VanTilburg:2015oza, Hees:2016gop, Leefer:2016xfu, Savalle:2020vgz}
and the laser interferometers \cite{Arvanitaki:2014faa, Stadnik:2015xbn, Pierce:2018xmy, Morisaki:2018htj, Grote:2019uvn, Morisaki:2020gui}.  
Searches for vector field dark matter with data from advanced Laser Interferometer Gravitational-Wave Observatory (aLIGO) \cite{LIGOScientific:2014pky} have already been carried out \cite{Guo:2019ker, LIGOScientific:2021ffg}.
The scalar wave can also manifest itself as the violation of the equivalence principle \cite{Damour:2010rp, Williams:2012nc, Wagner:2012ui, MICROSCOPE:2022doy}
and the fifth force \cite{Adelberger:2003zx}.
Since the amplitude of the scalar wave increases toward higher redshift,
the scalar dark matter also affects the abundance of light elements
produced by the big bang nucleosynthesis
and the temperature anisotropies of the cosmic microwave background \cite{Stadnik:2015kia}.

If the mass is in the range $(10^{-13}~{\rm eV}, 10^{-11}~{\rm eV})$,
the frequency of the signal falls in the range relevant to the
ground laser interferometers.
In \cite{Arvanitaki:2014faa, Morisaki:2018htj}, it was suggested that 
the laser interferometers can probe the scalar field dark matter by measuring
the motion of mirrors caused by the scalar field.
When the scalar field interacts with elementary particles,
the mass of the mirror varies in responding to the value of the scalar field
at the mirror's location. 
As a result, the mirror under the influence of the scalar field
accelerates toward a direction along which the mirror minimizes 
its energy (= mass).
Since the mass of an atom dominantly originates from the sector
of the strong interaction,
the scalar force that accelerates the mirror is most sensitive to
the coupling to the gluon fields.

In the non-relativisitc limit of the mirror's motion,
the force appears only when the scalar field $\phi$
varies in space, which means that the scalar force is parallel to 
${\bm \nabla}\phi$.
Such a force causes the modulation of the round-trip time of light between 
the front and the end mirrors, which eventually leads to the detectable signal
in the interferometers.
There are two physical effects that contribute to such modulation
of the round-trip time:
i) finiteness of speed of light which produces signal even if
the front and the end mirrors are perfectly synchronized (denoted
by $\delta t_1$ in \cite{Morisaki:2018htj})
and ii) difference of the displacement between the front and the
end mirrors due to spatial variation of the scalar field 
(denoted by $\delta t_2$ in \cite{Morisaki:2018htj}).
For aLIGO, the former/(latter) becomes dominant if the frequency of the scalar field oscillation is larger/(smaller) than $20~{\rm Hz}$.

The papers \cite{Arvanitaki:2014faa, Morisaki:2018htj} give expected 
upper limit on the magnitude of interaction between the scalar field dark matter
and elementary particles for the laser interferometers such as aLIGO.
To the best of our knowledge, there is no study which derived the upper limit 
by confronting the predicted signal with the real data obtained by the currently
operating laser interferometers.
In this paper, we report the result of our analysis using the real data of two LIGO
detectors taken during their third observation run.
In the analysis, we take into account the stochastic nature of the scalar field,
velocity dispersion of dark matter, and the anisotropic distribution
of velocity of dark matter due to rotation of the solar system in the Milky Way Galaxy. 
Formulation to compute the signal that includes the above effects and
can be readily compared with the data released
by LVK Collaboration is developed in Sec.~\ref{formulation}.
The upper limits based on our analysis are given in Sec.~\ref{result}.
 
Before closing this section, it should be mentioned that the scalar field
not only exerts the force but also changes the size of the bodies
such as beam splitters and mirrors.
The latter phenomenon occurs due to the variation of the atomic size ($\simeq$ Bohr radius)
which depends on the fine structure constant and the electron mass.
It was recognized in \cite{Grote:2019uvn} that the time variation of
the beam splitter produces detectable signal which is more prominent in GEO600 interferometer
than aLIGO.
The reason why GEO600 is more powerful than aLIGO is because GEO600 has a direct sensitivity to the signal, while in aLIGO, the strength of the signal is attenuated by a factor of arm cavity finesse.
Thus, GEO600 is suited for probing the couplings of the scalar field to the photons and electrons
and plays a role complementary to aLIGO which is sensitive to the coupling to the gluons.
In \cite{Vermeulen:2021epa}, the signal was searched for in the data of GEO600
and the upper limits on the couplings to photons and electrons were derived.

\section{Formulation of the signal}
\label{formulation}
In this section, we first give the Lagrangian of the model in which 
the dark matter is a classically
oscillating massive scalar field weakly interacting with elementary particles such as
electrons, quarks, photons, and gluons, and give a brief overview
of how dark matter yields signal in the laser interferometers.
We then develop the formulation which enables to directly compare the predicted signal
with public data provided by the LVK collaborations.

\subsection{Model considered in this paper}
The mathematical framework of the model we consider is given in \cite{Damour:2010rp}.
In this model, the scalar field $\phi$ which comprises all dark matter 
has interactions with elementary particles.
The Lagrangian density for the scalar field is given by
\begin{align}
{\cal L}&=-\frac{1}{2}{(\partial \phi)}^2-\frac{m^2}{2}\phi^2+{\cal L}_{\rm int}, \\
&{\cal L}_{\rm int}=-\kappa \phi \bigg[ \frac{d_e}{4e^2} F_{\mu \nu}F^{\mu \nu}-\frac{d_g\beta_3}{2g_3}
G_{\mu \nu}^A G^{A\mu \nu}
-\sum_{i=e,u,d} (d_{m_i}+\gamma_{m_i} d_g ) m_i {\bar \psi_i}\psi_i
\bigg].
\end{align}
Here $\kappa \equiv \frac{\sqrt{4\pi}}{M_{\rm pl}}$,
and $M_{\rm pl}=1/\sqrt{G}\simeq 1.22\times 10^{19}~{\rm GeV}$ is the Planck mass.
The first, second, and third term in ${\cal L}_{\rm int}$ represents interaction
between $\phi$ and electromagnetic field, gluon fields, and fermions including
electrons, up quarks, and down quarks, respectively.
Magnitude of each interaction is parametrized by dimensionless constant $d_e, d_g, d_{m_i}$
which we treat as free parameters.
One interesting example in which these five parameters are parametrized by
a single parameter as $d_g=d_i,~d_e=0$ ($i=e,u,d$) is realized if  
the scalar field has the nonminimal coupling to
gravity in the Jordan frame. 
Description of this model is given in the Appendix.
$\beta_3$ is the QCD beta function and $\gamma_{m_i}$ is the anomalous dimension.
The interaction Lagrangian given above is defined at some high energy scale.
We can translate the effect of ${\cal L}_{\rm int}$ into equivalent but more physically relevant 
representation by considering the renormalization-group flow:
\begin{align}
&\alpha (\varphi)=(1+d_e \varphi)\alpha, \label{alpha-phi}\\
&\Lambda_3 (\varphi)=(1+d_g \varphi) \Lambda_3, \label{Lambda3-phi} \\
&m_e (\varphi)=(1+d_{m_e} \varphi )m_e, \label{me-phi} \\
&m_{u,d} (\varphi)=(1+d_{m_u, m_d} \varphi)m_{u,d}(\Lambda_3). \label{mud-phi}
\end{align}
Here, $\varphi \equiv \kappa \phi$ is the scalar field normalized by the Planck scale.
$\alpha$ is the fine-structure constant, $\Lambda_3$ is the QCD mass scale,
and $m_{u,d}$ are the quark masses defined at the QCD mass scale.
Due to the presence of $\varphi$, these physical constants deviate from those in the absence of $\varphi$.
We assume that the mass $m$ is in the range $10^{-13}~{\rm eV} \lesssim m \lesssim 10^{-11}~{\rm eV}$.
In this case, the $\phi$ field as dark matter behaves as a classically oscillating wave
since the corresponding occupation number is huge (e.g., \cite{Hui:2021tkt}).
In Fourier domain, frequencies are sharply localised at $f \simeq \frac{m}{2\pi}$ within
the width $\Delta f/f = \frac{v_*^2}{2} \simeq 10^{-6}$, where
$v_*=220~{\rm km/s}$ is the velocity dispersion of dark matter.
Thus, oscillations of $\phi$ produces almost monochromatic time variation of 
the physical constants listed in Eqs.~(\ref{alpha-phi})-(\ref{mud-phi}).
The frequency range corresponding to the mass range $10^{-13}~{\rm eV} \lesssim m \lesssim 10^{-11}~{\rm eV}$ is $10~{\rm Hz} \lesssim \frac{m}{2\pi} \lesssim 10^3~{\rm Hz}$ which falls in the
frequency band of the ground-based laser interferometers.

As it is argued in \cite{Damour:2010rp}, coupling to the fermion kinetic term can be eliminated 
by the field redefinition.
To see this, let us suppose that the $\phi$ couples to the kinetic term of the fermion as
\be
{\cal L}=-e^{f(\phi)} {\bar \psi} \gamma^\mu \partial_\mu \psi-\frac{1}{2}e^{f(\phi)}
\partial_\mu f {\bar \psi}\gamma^\mu \psi,
\ee
where we leave $f(\phi)$ unspecified to keep generality of our argument.
The second term is necessary to make the kinetic term be Hermitian.
Then, by changing the variable as $\psi \to e^{-\frac{1}{2} f}\psi$, 
the kinetic term in terms of the new field becomes 
\be
{\cal L}=-{\bar \psi} \gamma^\mu \partial_\mu \psi,
\ee
and the coupling $f(\phi)$ has been completely absorbed into the field redefinition.

Since mass of an atom depends on physical constants,
it depends on $\varphi$ through Eqs.~(\ref{alpha-phi})-(\ref{mud-phi}).
Denoting the mass of an atom by $m_A (\varphi)$,
$m_A$ in the small $\varphi$ limit is approximately written as \cite{Damour:2010rp}
\begin{align}
\label{alpha_A}
\alpha_A &\equiv \frac{d\ln m_A(\varphi)}{d\varphi} \nonumber \\
&\simeq d_g^*+\bigg[-\frac{0.036}{A^{\frac{1}{3}}}-0.02 \frac{{(A-2Z)}^2}{A^2}-1.4\times 10^{-4} \frac{Z(Z-1)}{A^\frac{4}{3}} \bigg] (d_{\hat m}-d_g) \nonumber \\
&~~~-2.7\times 10^{-4} \frac{A-2Z}{A} (d_{m_e}-d_g)+
\left( -4.1 \frac{A-2Z}{A}+7.7 \frac{Z(Z-1)}{A^\frac{4}{3}} \right) \times 10^{-4}d_e, 
\end{align}
and $d_g^*$ is defined by
\be
d_g^* = d_g+0.093 (d_{\hat m}-d_g)+2.7\times 10^{-4} (d_{m_e}-d_g)+2.7\times 10^{-4}d_e,
\ee
where $A, Z$ is the mass number and the atomic number, respectively,
and 
\be
d_{\hat m}\equiv \frac{d_{m_u}m_u+d_{m_d}m_d}{m_u+m_d}.
\ee
As it is clear, $d_g^*$ represents a part independent of the atomic species 
and the remaining terms are dependent on the atomic species.
Thus, any model belonging to a class satisfying $d_{\hat m}=d_{m_e}=d_g, ~d_e=0$ 
does not violate the weak equivalence principle.
A non-minimal coupling model described in the Appendix belongs to this class.

The time variation of the mass of atoms due to the oscillations of $\phi$ 
causes the time variation 
of the mass $M(\varphi)$ of a macroscopic object.
The equation of motion of the macroscopic object under the influence of $\varphi$
is given by (e.g., \cite{Morisaki:2018htj})
\be
\label{phi-eom}
\frac{d^2 {\bm x}}{dt^2} \simeq -\langle \alpha_A \rangle {\bm \nabla}\varphi,
\ee
where $\langle \alpha_A \rangle$ is $\alpha_A$ averaged over different types of
molecules constituting the macroscopic object.
The mirrors installed in the interferometers are purely made of silica (=${\rm SiO_2}$).
In this case, we have
\be
\label{alpha-sio2}
\alpha_{SiO_2} \simeq
d_g+0.083(d_{\hat m}-d_g)+2.7\times 10^{-4} (d_{m_e}-d_g)+3.1\times 10^{-3}d_e.
\ee
It is seen that the mirrors are mostly sensitive to the coupling to the gluon
fields $d_g$ and the couplings to electrons and photons are suppressed.
This is because the mass of nucleons is mainly determined by the strong interaction.

\subsection{Scalar field in the Milky Way Galaxy}
\subsubsection{Scalar field in Galaxy's center-of-mass frame}
Inside the Milky Way Galaxy, dark matter is virialized.
The ocsillating scalar field at the location of the solar system 
in the galaxy's center-of-mass frame may be modeled as stochastic wave with stationary and isotropic velocity distribution
with velocity dispersion $v_* \simeq 220~{\rm km/s}$.
In this coordinate system, 
the scalar dark matter may be expressed as the superposition of 
plane waves as
\be
\label{sto-phi}
\phi (t, {\bm x})=\int d {\hat {\bm \Omega}} \int_{-\infty}^\infty df~
e^{-2\pi i f (t-\eta {\hat {\bm \Omega}} \cdot {\bm x})} u (f,{\hat {\bm \Omega}}).
\ee
Here, $\eta$ is defined by
\be
\eta = \sqrt{1-\frac{\mu^2}{f^2}},~~~~~~~~~\left(\mu=\frac{m}{2\pi} \right),
\ee
and ${\hat {\bm \Omega}}$ is the unit vector representing the propagation direction of the
plane wave $u(f,{\hat {\bm \Omega}})$.
From the reality of $\phi$, we have $u^* (f,{\hat {\bm \Omega}})=u(-f,{\hat {\bm \Omega}})$.
From the stationarity and isotropy condition, we assume that the ensemble average
of the stochastic variable $u(f,{\hat {\bm \Omega}})$ can be written as
\be
\label{def-correlation-phi}
\langle u^* (f,{\hat {\bm \Omega}}) u(f',{\hat {\bm \Omega'}}) \rangle
=\delta (f-f') \frac{1}{4\pi} \delta ({\hat {\bm \Omega}},{\hat {\bm \Omega'}})
\times \frac{1}{2}P_\phi (f).
\ee
Here, $P_\phi (f)$ is the powerspectrum of $\phi$ whose shape may be determined
by comparing the energy density of the scalar field evaluated in
the wave picture with that in the particle picture in the following way.
Let us first evaluate the energy density in the wave picture.
The energy density of the scalar field is given by
\be
\rho_\phi =\frac{1}{2}{\dot \phi}^2+\frac{1}{2} {({\bm \nabla}\phi )}^2+\frac{m^2}{2}\phi^2.
\ee
Using Eq.~(\ref{sto-phi}), the ensemble average of $\rho_\phi$ becomes
\be
\label{Sphi-rho}
\langle \rho_\phi \rangle =\int_\mu^\infty df~{(2\pi f)}^2 P_\phi (f).
\ee

\begin{figure}[t]
  \begin{center}
    \includegraphics[clip,width=10.0cm]{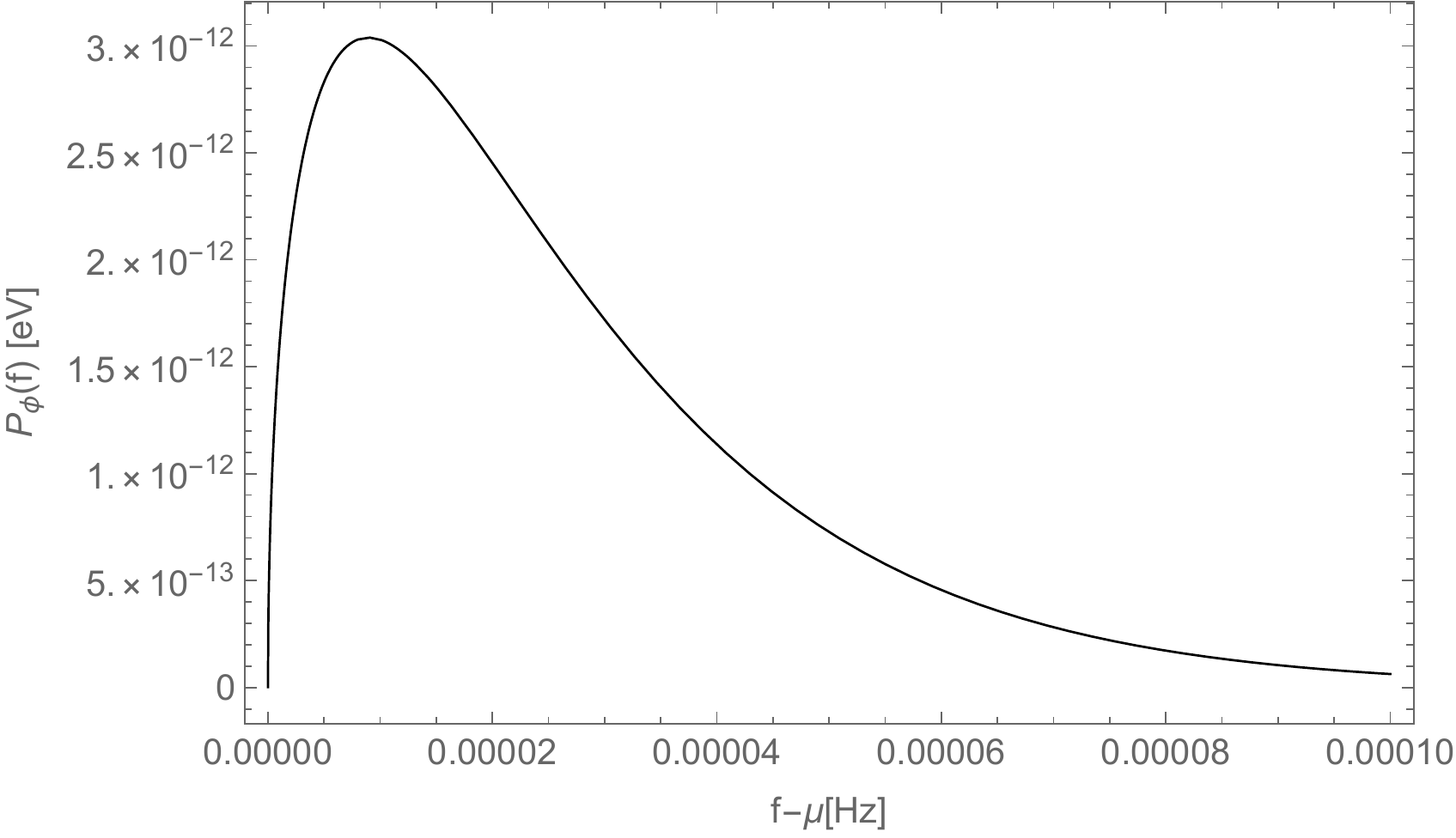}
    \caption{Plot of $P_\phi (f)$ for $\mu=100~{\rm Hz}$, $v_*=220 {\rm km/s}$.}
    \label{fig.S-phi}
  \end{center}
\end{figure}

Next, we evaluate the same quantity in the particle picture.
In this picture, we interpret the stochastic scalar field as a collection of 
randomly moving particles with mass $m$, for which we can introduce a notion 
of the distribution function $g({\bm v})$ in the velocity space.
As it is usually done, 
we assume that the distribution of the particles obeys the Boltzmann distribution as
\be
g({\bm v})=\frac{3^\frac{3}{2} \rho_{\rm loc}}{{(2\pi)}^{3/2}m v_*^3} 
\exp \left( -\frac{3v^2}{2v_*^2} \right).
\ee
Here $v=|{\bm v}|$, and $v_*$ is the velocity dispersion of dark matter.
The normalization constant is fixed by requiring that the integral of 
$g({\bm v})$ coincides with the local
number density of dark matter particles;
\be
\int d{\bm v}~g({\bm v})=\frac{\rho_{\rm loc}}{m}.
\ee
Therefore, we have
\be
\label{local-dm}
\int d{\bm v} \frac{3^\frac{3}{2} \rho_{\rm loc}}{{(2\pi)}^{3/2} v_*^3}
\exp \left( -\frac{3v^2}{2v_*^2} \right)=\rho_{\rm loc}.
\ee
In order to compare, we change the integration variable from $v$ to $f$ by using the relation
\be
\label{f-v}
f^2=\mu^2 (1+v^2).
\ee
In terms of $f$, Eq.~(\ref{local-dm}) becomes
\be
\int_\mu^\infty df~{\left( \frac{3}{2} \right)}^\frac{3}{2}
\frac{4\rho_{\rm loc}}{\sqrt{\pi} \mu^3 v_*^3} f^2\eta 
\exp \left( -\frac{3f^2 \eta^2}{2v_*^2\mu^2} \right) =\rho_{\rm loc}
\ee
By identifying this equation with Eq.~(\ref{Sphi-rho}), we can express $P_\phi (f)$ in terms of the 
Boltzmann distribution function as
\be
\label{B-S-phi}
P_\phi (f)={\left( \frac{3}{2} \right)}^\frac{3}{2}
\frac{4\rho_{\rm loc}}{\pi^{5/2} \mu^3 v_*^3} \eta 
\exp \left( -\frac{3f^2 \eta^2}{2v_*^2\mu^2} \right).
\ee
Fig.~\ref{fig.S-phi} shows a plot of $P_\phi (f)$ for $\mu =100{\rm Hz}, v_*=220 {\rm km/s}$.

\subsubsection{Scalar field in the Earth's rest frame}
Since observations by the terrestrial detectors are conducted on the Earth,
we need to translate the scalar field represented in the galaxy's center-of-mass frame
into the one in the Earth's rest frame.
Although the Earth, which is orbiting in the Milky Way Galaxy, is accelerating 
with respect to the center-of-mass frame of the Galaxy,
we ignore the effect of the acceleration and make an approximation that the Earth
is moving with a constant velocity ${\bm v_0}$ in the Galaxy.
We also ignore the Earth's orbital motion around the Sun since the orbital velocity
($\simeq 30~{\rm km/s}$) is much smaller than $|{\bm v_0}| \simeq 200~{\rm km/s}$.
To be definite, we put {\it prime} on coordinates and relevant quantities when they are
defined in the Galaxy's rest frame. For instance, spacetime coordinates are written as
$(t',{\bm x'})$, and frequency and the propagation direction are written
as $(f', {\hat \Omega'})$.
On the other hand, quantities without prime are defined in the Earth's rest frame.
Based on our assumptions, both frames are inertial systems and quantities in different
frames are related by the Lorentz transformation.
Frequency $f$ and the wavenumber vector ${\bm k}$ transform as
\begin{align}
&f'=\gamma \left( f+\frac{{\bm v_0}\cdot {\bm k}}{2\pi} \right),~~~~~
\gamma \equiv \frac{1}{\sqrt{1-{\bm v_0}^2}} \\
&{\bm k'}=\gamma \left( {\bm k}+2\pi f {\bm v_0} \right).
\end{align}
Since motions are non-relativistic (i.e., 
$|{\bm v_0}| \simeq v_* ={\cal O}(10^{-3}) \ll 1$),
we retain only the relevant lowest order terms in ${\bm v_0}$ or ${\bm k}/m$ 
in the above relations.
After some manipulations, we obtain
\begin{align}
\label{L-transform}
f'&=f+\frac{\mu}{2}{\bm v_0}^2+\mu \eta {\bm v_0}\cdot {\hat {\bm \Omega}}, \\
k'&=\Big| {\hat {\bm \Omega}}+\frac{\bm v_0}{\eta} \Big| k, \\
{\hat {\bm \Omega'}}&={\Big| {\hat {\bm \Omega}}+\frac{\bm v_0}{\eta} \Big|}^{-1}
\left( {\hat {\bm \Omega}}+\frac{\bm v_0}{\eta} \right).
\end{align}

Since $\phi$ itself is Lorentz-invariant, Eq.~(\ref{sto-phi}), which is expressed in 
terms of the variables of the Galaxy's rest frame, may be directly used to represent
$\phi$ in the Earth's rest frame.
We note that a combination $kdf d{\hat \Omega}$,
where $k$ is the wavenumber,
is Lorentz-invariant:
\be
k' df' d{\hat {\bm \Omega'}}=kdf d{\hat {\bm \Omega}}.
\ee
By exploiting this relation,
Eq.~(\ref{sto-phi}) may be written as
\be
\phi (t,{\bm x}) =\int d{\hat {\bm \Omega}}df ~\frac{k}{k'} 
e^{-2\pi i f (t-\eta {\hat {\bm \Omega}} \cdot {\bm x})} u (f',{\hat {\bm \Omega'}}).
\ee
Note that $(t,{\bm x})$ are coordinates in the Earth's rest frame.
Plugging the transformation rules for $f', k',$ and ${\bm \Omega'}$, we have
\be
\label{sto-phi-Earth}
\phi (t,{\bm x})=\int d{\hat {\bm \Omega}}df ~
e^{-2\pi i f (t-\eta {\hat {\bm \Omega}} \cdot {\bm x})} 
{\Big| {\hat {\bm \Omega}}+\frac{\bm v_0}{\eta} \Big|}^{-1}
u \left( f+\frac{\mu}{2}{\bm v_0}^2+\mu \eta {\bm v_0}\cdot {\hat {\bm \Omega}},
\frac{\eta {\hat {\bm \Omega}}+{\bm v_0} }{ | \eta {\hat {\bm \Omega}}+{\bm v_0}|} \right).
\ee
For notational convenience, we introduce a new function ${\bar u}(f,{\hat {\bm \Omega}})$
by
\be
\label{def-bar-u}
{\bar u}(f,{\hat {\bm \Omega}})={\Big| {\hat {\bm \Omega}}+\frac{\bm v_0}{\eta} \Big|}^{-1}
u \left( f+\frac{\mu}{2}{\bm v_0}^2+\mu \eta {\bm v_0}\cdot {\hat {\bm \Omega}},
\frac{\eta {\hat {\bm \Omega}}+{\bm v_0} }{ | \eta {\hat {\bm \Omega}}+{\bm v_0}|} \right),
\ee
so that $\phi$ may be written as
\be
\label{sto-phi-Earth-2}
\phi (t,{\bm x})=\int d{\hat {\bm \Omega}}df ~
e^{-2\pi i f (t-\eta {\hat {\bm \Omega}} \cdot {\bm x})} 
{\bar u}(f,{\hat {\bm \Omega}}).
\ee
This is the expression of $\phi$ that is used in the rest of this paper.

\subsection{Detector's signal of the scalar field}
A mirror obeys the equation of motion given by Eq.~(\ref{phi-eom}) and
undergoes oscillations caused by the scalar force ${\bm \nabla}\varphi$.
Response of a mirror by the action of the scalar field was computed in \cite{Morisaki:2018htj}.
For the scalar field given by Eq.~(\ref{sto-phi-Earth-2}), mirror's motion is written as
\be
\delta {\bm x} (t)=-\kappa \alpha_{SiO_2} \int d{\hat {\bm \Omega}} \int_{-\infty}^\infty
\frac{df}{2\pi i f}\eta e^{-2\pi i f (t-\eta {\hat {\bm \Omega}} \cdot {\bm x})} 
{\bar u} (f,{\hat {\bm \Omega}})
{\hat {\bm \Omega}}.
\ee
Then, using the result in \cite{Morisaki:2018htj}, 
the signal $s(t)$ of the Michelson-type interferometer located at ${\bm x}$, 
which is the phase difference of the lasers propagating along each arm,
can be computed. The result is given by
\be
s(t)=\int d{\hat {\bm \Omega}} \int_{-\infty}^\infty df~F({\hat {\bm \Omega}},f) 
e^{-2\pi i f (t-\eta {\hat {\bm \Omega}} \cdot {\bm x})} 
{\bar u}(f,{\hat {\bm \Omega}}),
\ee
where the detector pattern function is given by
\be
F({\hat {\bm \Omega}},f)=-\frac{\kappa \alpha_{SiO_2}}{\pi iLf} \left( {\hat {\bm m}}\cdot {\hat {\bm \Omega}}-
{\hat {\bm n}}\cdot {\hat {\bm \Omega}} \right) \eta \sin^2 (\pi f L)
-\kappa \alpha_{SiO_2} \left( {( {\hat {\bm m}}\cdot {\hat {\bm \Omega}} )}^2-
{( {\hat {\bm n}}\cdot {\hat {\bm \Omega}} )}^2 \right) \eta^2. 
\ee
The first/(second) term corresponds to $\delta t_1$/$(\delta t_2)$ defined in \cite{Morisaki:2018htj}, respectively. 
This function satisfies $F^*({\hat {\bm \Omega}},f)=F(-{\hat {\bm \Omega}},f)$.

The Fourier transform of the signal is given by
\be
\label{Fourier-signal}
{\tilde s}(f)=\int d{\hat {\bm \Omega}} ~F({\hat {\bm \Omega}},f) 
e^{2\pi i f \eta {\hat {\bm \Omega}} \cdot {\bm x}} {\bar u}(f,{\hat {\bm \Omega}})
\ee

\subsubsection{Two-detectors correlation}
In the observation, the detector's output $S(t)$ consists of the signal $s$ (if it exists)
and the noise $n$ as
\be
S(t)=s(t)+n(t).
\ee
Since $\varphi$ is a stochastic variable, the signal $s(t)$ also behaves in a stochastic manner.
Practically, it is difficult to distinguish the signal from the noise.
As it is usually done, we thus consider correlation of the outputs between two detectors 
in order to extract the signal from the noise.

Using Eqs.~(\ref{Fourier-signal}), (\ref{def-correlation-phi}), and (\ref{def-bar-u}), 
we can compute the cross correlation of the outputs between the detector 1 and the detector 2 as
\be
\langle {\tilde S_1^*}(f) {\tilde S_2} (f')\rangle
=\int \frac{d{\hat {\bm \Omega}}}{4\pi} F_1^* ({\hat {\bm \Omega}},f) F_2 ({\hat {\bm \Omega}},f)
e^{2\pi i f \eta {\hat {\bm \Omega}} \cdot {\bm \Delta x}}\times \frac{1}{2} \delta (f-f') P_\phi 
(f+\frac{\mu}{2}{\bm v_0}^2+\mu \eta {\bm v_0}\cdot {\hat {\bm \Omega}}).
\ee
Here, it has been assumed that the correlation of the noise between the two detectors is zero.
${\bm \Delta x}={\bm x_2}-{\bm x_1}$ is the separation between the two detectors.
Defining the overlap function $\Gamma_\phi (f)$ in the absence of the Earth's motion as
\be
\label{Gamma-phi}
\Gamma_\phi (f) \equiv \int \frac{d{\hat {\bm \Omega}}}{4\pi} F_1^* ({\hat {\bm \Omega}},f) 
F_2 ({\hat {\bm \Omega}},f)
e^{2\pi i f \eta {\hat {\bm \Omega}} \cdot {\bm \Delta x}},
\ee
the cross correlation becomes
\be
\langle {\tilde S_1^*}(f) {\tilde S_2} (f')\rangle = \frac{\delta (f-f')}{2}\Gamma_\phi (f) P_\phi (f)
\varepsilon (f).
\ee
Here $\varepsilon (f)$ defined by
\be
\varepsilon (f) \equiv \frac{1}{\Gamma_\phi (f) P_\phi (f)}
\int \frac{d{\hat {\bm \Omega}}}{4\pi} F_1^* ({\hat {\bm \Omega}},f) F_2 ({\hat {\bm \Omega}},f)
e^{2\pi i f \eta {\hat {\bm \Omega}} \cdot {\bm \Delta x}} 
P_\phi 
(f+\frac{\mu}{2}{\bm v_0}^2+\mu \eta {\bm v_0}\cdot {\hat {\bm \Omega}})
\ee
quantifies the change of the correlation signal caused by the motion
of the solar system with respect to the rest frame of the Milky Way Galaxy.
Notice that $\varepsilon (f)$ is a complex number.

\begin{figure}[t]
  \begin{center}
    \includegraphics[clip,width=12.0cm]{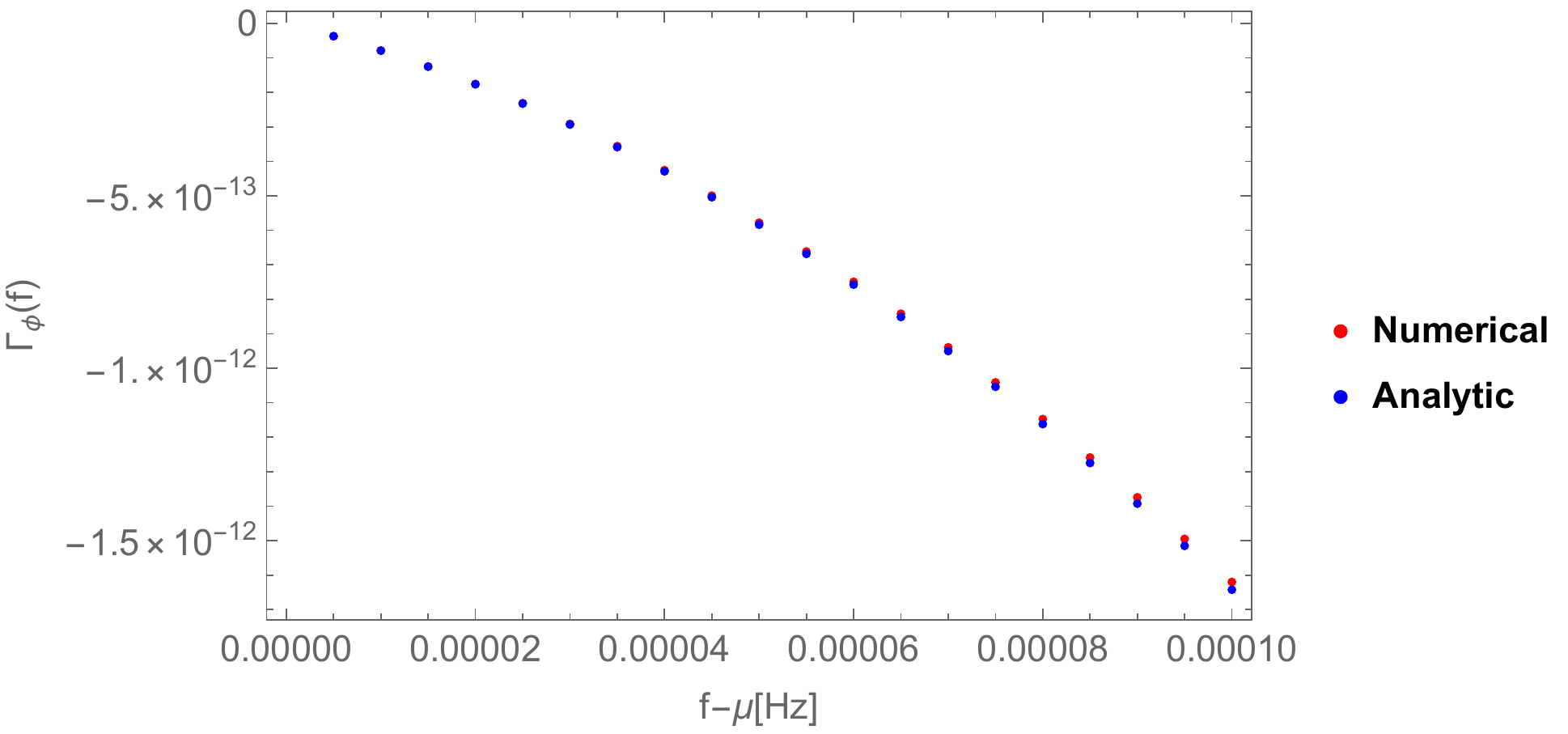}
    \caption{Plot of $\Gamma_\phi (f)$ for $\mu=100~{\rm Hz}$, $\kappa \alpha_{SiO_2}=1$.}
    \label{fig.Gamma-phi}
  \end{center}
\end{figure}

From the relation $F^*({\hat {\bm \Omega}},f)=F(-{\hat {\bm \Omega}},f)$, 
we can verify that $\Gamma_\phi (f)$
is a real function.
For LIGO interferometers, we have $fL \ll 1$ and $2\pi f\eta |{\hat {\bm \Omega}}\cdot {\bm \Delta x}| \ll 1$.
Thus, it is a good approximation to ignore the phase proportional to ${\bm \Delta x}$ in Eq.~(\ref{Gamma-phi}).
With this approximation, we can perform the angular integration analytically;
\be
\label{approx-Gamma-phi}
\Gamma_\phi (f) \approx {( \kappa \alpha_{SiO_2} \eta )}^2 \left( a_1 {( f L)}^2 +
a_2 \eta^2 \right).
\ee
where $a_1, a_2$ are defined by
\begin{align}
a_1 &= \frac{\pi^2}{3} 
({\hat m_1}-{\hat n_1})\cdot ({\hat m_2}-{\hat n_2}), \\
a_2 &=\frac{2}{15}\big[ \left( ({\hat m_1}+{\hat n_1})\cdot {\hat m_2} \right)
\left( ({\hat m_1}-{\hat n_1})\cdot {\hat m_2} \right)
-({\hat m_2} \leftrightarrow {\hat n_2})\big].
\end{align}
Fig.~\ref{fig.Gamma-phi} shows numerically computed $\Gamma_\phi (f)$
and the approximate one (\ref{approx-Gamma-phi}) for $\kappa \alpha_{SiO_2} =1$ \footnote{
Numerical values of the vectors ${\hat m}, {\hat n}$ for the existing detectors can be found at
\url{https://lscsoft.docs.ligo.org/lalsuite/lal/_l_a_l_detectors_8h_source.html}}.
Clearly, the approximate formula nicely reproduces the exact one.

\begin{figure}[t]
  \begin{center}
    \includegraphics[clip,width=10.0cm]{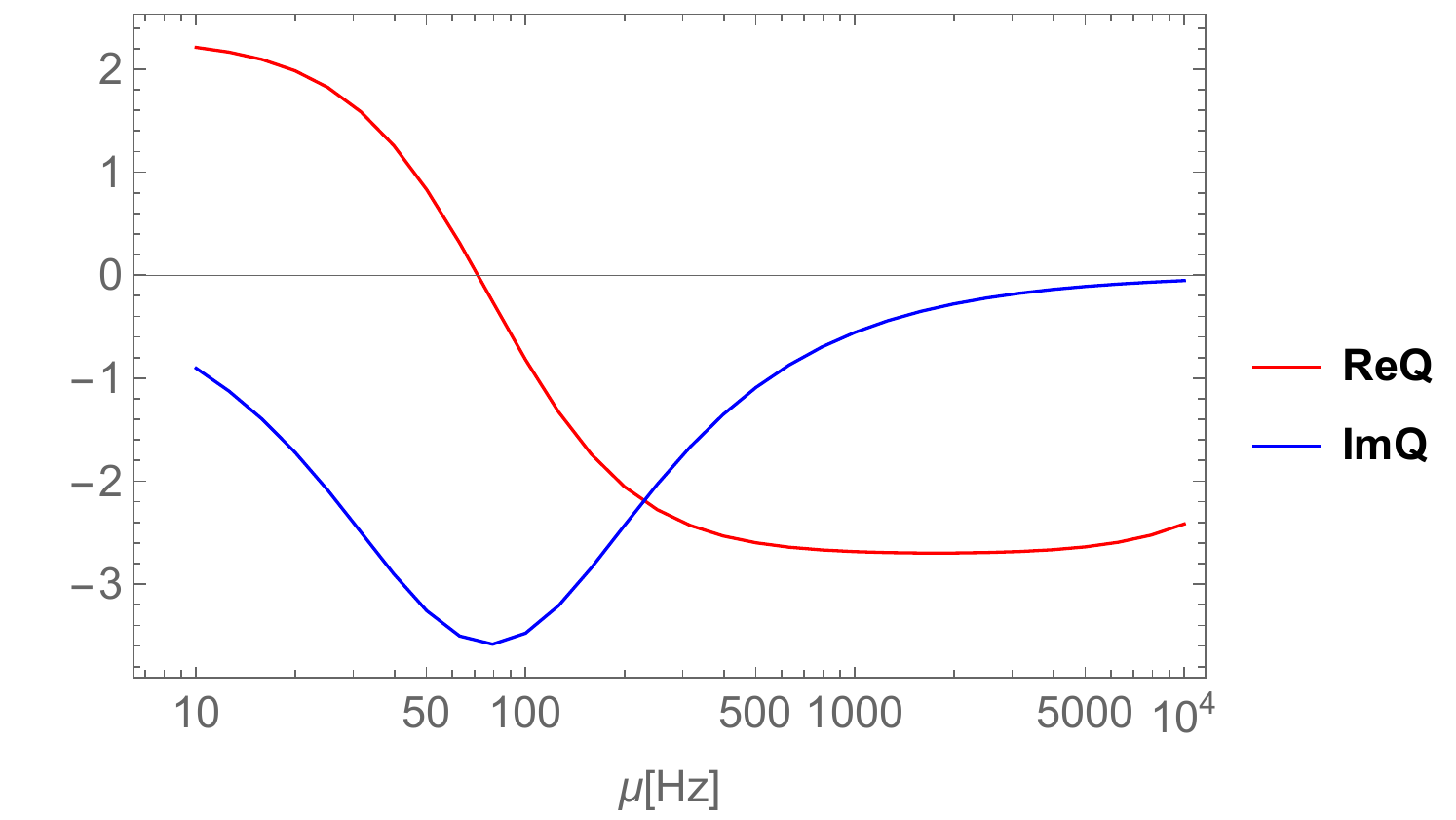}
    \caption{Plot of ${\rm Re} ~Q(\mu)$ and ${\rm Im} ~Q(\mu)$.}
    \label{fig.Q}
  \end{center}
\end{figure}

Our formulation expanded up to this point assumes infinitely long observation time $T$. 
However, in reality, the processed public data which we can use, 
is obtained for $T=192~{\rm s}$.
Thus, we need to relate the real observable with what we have computed for $T \to \infty$.
First, suppose that the data $S(t)$ was taken during the period $(-T/2, T/2)$.
We define the Fourier transformation of the data $S(t)$ as
\be
{\tilde S}_{\rm T} (f)=\int_{-\frac{T}{2}}^{\frac{T}{2}}~dt~e^{2\pi i ft }S(t) W_T(t),
\ee
where $W_T(t)$ is the Hann Window function which is used in LVK Collaboration.
On the left-hand side, we have put the subscript ``${\rm T}$" to emphasize that it is 
directly computed from data.
Meanwhile, the original Fourier transformation is defined by
\be
{\tilde S} (f)=\int_{-\infty}^{\infty}~dt~e^{2\pi i ft }S(t).
\ee
From these equations, we find that ${\tilde S}_{\rm T} (f)$
and ${\tilde S} (f)$ are related as
\be
{\tilde S}_{\rm T} (f)=\int_{-\infty}^\infty df'~
{\tilde S}(f') {\tilde W}_T(f-f'),
\ee
where ${\tilde W}_T(f)$ is defined by
\be
{\tilde W}_T(f)=\int_{-\frac{T}{2}}^{\frac{T}{2}}~dt~
e^{2\pi i f t}W_T(t) =\sqrt{\frac{8}{3}} \frac{\sin (\pi f T)}{2\pi f (1-f^2 T^2)},
\ee
and in the last equation we have given an explicit expression for the Hann Window function.
Using this relation, we can derive the cross-correlation of ${\tilde S}_{\rm T} (f)$
in terms of $P_\phi (f)$ and $\Gamma_\phi (f)$ as
\be
\langle {\tilde S}_{1,\rm T}^* (f) {\tilde S}_{2,\rm T} (f) \rangle
=\frac{1}{2} \int_{-\infty}^\infty {|{\tilde W}_T(f-f')|}^2
\Gamma_\phi (f') P_\phi (f') \varepsilon (f') df'.
\ee
To evaluate this integral approximately, we note $P_\phi (f')$ is sharply peaked at 
$\mu \le f' \lesssim \mu+\mu v_*^2 $.
Thus, the integration is contributed only from this short interval.
Since $\mu v_*^2 \ll 1/T $, the Window function remains almost constant in this interval. 
Then, it is a good approximation to replace $f'$ with $\mu$ in ${\tilde W}_T (f-f')$
and pull it out of the integration;
\begin{align}
\label{approx-cc}
\langle {\tilde S}_{1,\rm T}^* (f) {\tilde S}_{2,\rm T} (f) \rangle
&\approx \frac{1}{2} {|{\tilde W}_T (\mu -f)|}^2
\left( \int_\mu^\infty 
\Gamma_\phi (f') P_\phi (f') df' \right) Q(\mu).
\end{align}
Here $Q(\mu)$ defined by
\be
Q(\mu) \equiv \frac{\int_\mu^\infty \Gamma_\phi (f) P_\phi (f) \varepsilon (f) 
df}{\int_\mu^\infty \Gamma_\phi (f) P_\phi (f) df}
\ee
quantifies the effect caused by the motion of the solar system.
Because $\varepsilon (f)$ is a complex number, $Q(\mu)$ is also a complex number.
Fig.~\ref{fig.Q} shows a plot of ${\rm Re} ~Q(\mu)$ and ${\rm Im} ~Q(\mu)$.
In making this plot, it is assumed that dark matter wind comes from the direction
specified by $(\ell, b)=(270^\circ, 0)$ in the galactic coordinates
which correspond to $(\alpha, \delta)=(3.70231, -0.81267)$ in the equilateral coordinates
and the signal is averaged over $\alpha$ to take into account the Earth's daily rotation.
We find that the signal is suppressed at $\mu \simeq 70~{\rm Hz}$ ($m \simeq 3\times 10^{-13}~{\rm eV}$)
whereas it is enhanced by a factor of $\sim 2$ at other frequencies.
We expect that the imaginary part of $Q$ may be used as additional observable
to increase the sensitivity of the search of the scalar dark matter by using the
laser interferometers.
Since only the real part of the cross correlation is publically available, 
we do not make use of ${\rm Im}~Q$ in our analysis.

It is possible to perform the integral $\int_\mu^\infty \Gamma_\phi P_\phi df$ analytically
to a good approximation. 
To this end, using Eqs.~(\ref{B-S-phi}) and (\ref{approx-Gamma-phi}),
we obtain
\be
\int_\mu^\infty 
\Gamma_\phi (f) P_\phi (f) df
\approx {(\kappa \alpha_{SiO_2})}^2 \int_\mu^\infty df~
\eta^2 \left( a_1 \mu^2 L^2+a_2 \eta^2 \right) \times
{\left( \frac{3}{2} \right)}^\frac{3}{2}
\frac{4\rho_{\rm loc} \eta}{\pi^{5/2} \mu^3 v_*^3} 
\exp \left( -\frac{3 \eta^2}{2v_*^2} \right).
\ee
where we have replaced $f$ which is not in $\eta$ with $\mu$.
By changing variable from $f$ to $v$ by the relation (\ref{f-v}), we have
\be
\int_\mu^\infty 
\Gamma_\phi (f') P_\phi (f') df'
\approx {(\kappa \alpha_{SiO_2})}^2 \frac{\rho_{\rm loc} v_*^2}{2\pi^2 \mu^2}
\left( 2a_1 \mu^2 L^2+\frac{10}{3} a_2 v_*^2 \right).
\ee
Thus, our final expression of the cross correlation becomes
\be
\label{final-cross-correlation}
\langle {\tilde S}_{1,\rm T}^* (f) {\tilde S}_{2,\rm T} (f) \rangle
\approx {(\kappa \alpha_{SiO_2})}^2 \frac{\rho_{\rm loc} v_*^2}{2\pi^2 \mu^2}
\left( a_1 \mu^2 L^2+\frac{5}{3} a_2 v_*^2 \right)
{|{\tilde W}_T (\mu -f)|}^2 Q(\mu).
\ee

The cross-correlation statistic ${\hat C}(f)$ given by the LVK collaboration is defined 
as a summation over discretized frequency bins with its width $1/T$;
\be
{\hat C}(f) =
\frac{1}{T^2 \Delta f} \sum_{i=0}^5 \frac{{\rm Re}[ {\tilde S_1^*}(f_i){\tilde S_2}(f_i)]}{\gamma_T (f_i) S_0 (f_i)},~~~~~(\Delta f=0.03125~{\rm Hz},~T=192~{\rm s})
\ee
where $f_i =f-\Delta f/2 +i/T$, $\gamma_T$ is the overlap reduction function for
gravitational waves, and $S_0 \equiv 3H_0^2/(10\pi^2 f^3)$.
In the presence of the scalar field dark matter, using Eq.~(\ref{final-cross-correlation}), 
the expectation value of ${\hat C}(f)$ becomes
\be
\label{model-C}
\langle {\hat C}(f) \rangle=
\frac{{(\kappa \alpha_{SiO_2})}^2 \rho_{\rm loc}v_*^2}{2\pi^2\mu^2 T^2 \Delta f} 
\left( a_1 \mu^2L^2+\frac{5}{3} a_2 v_*^2 \right) \Re Q(\mu) \sum_{i=0}^5
\frac{{|{\tilde W}_T (\mu-f_i)|}^2}{\gamma_T (f_i) S_0 (f_i)}.
\ee
This is the actual signal that we seek in the GW data provided by the LVK Collaboration.

\section{Results}
\label{result}
We search for the signal given by Eq.~(\ref{model-C}) in the data of the two LIGO 
detectors during their third observing run (O3).
The LVK Collaboration gives the cross correlation statistics which has been obtained
by computing ${\hat C}$ for each segment containing data of the duration $T$
and averaging them \footnote{Data is available at \url{https://dcc.ligo.org/cgi-bin/DocDB/ShowDocument?.submit=Identifier&docid=G2001287&version=5}}.  
Thus, the cross correlation statistics obeys the Gaussian distribution, 
and the likelihood becomes \cite{Mandic:2012pj}
\be
p({\hat C}|{\bm \Theta}) \propto \exp 
\left( -\sum_f \frac{{({\hat C}(f)-\langle {\hat C}(f)\rangle )}^2 }{2\sigma^2(f)} \right),
\ee
where ${\bm \Theta}=\{ \mu, \alpha_{Si O_2} \}$ is a set of parameters characterizing
the scalar field model.

In our analysis, we employ the Bayesian inference. 
We first fix $\mu$ and compute the posterior on $\alpha_{Si O_2}$ by adopting
two priors (uniform in $\alpha_{Si O_2}$ and uniform in $\ln \alpha_{Si O_2}$).
We then repeat this calculation by scanning the relevant range of $\mu$.
We did not find any conclusive evidence of a signal in data. 
As a result, we are able to place upper limit on $\alpha_{Si O_2}$ for various values of $\mu$.

\begin{figure}
\begin{center}
\begin{tabular}{cc}
\begin{minipage}[t]{0.5\hsize}
\includegraphics[width=7cm]{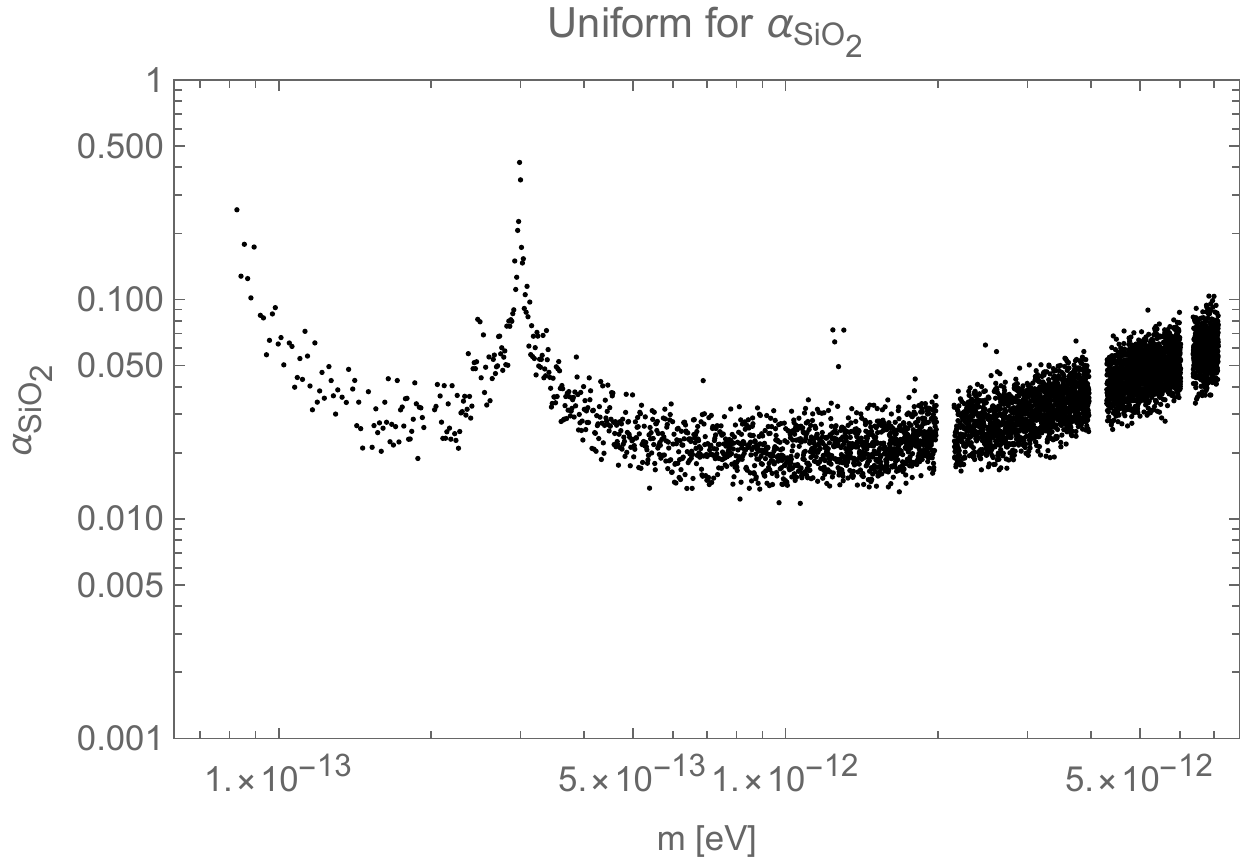}
\end{minipage}
\begin{minipage}[t]{0.5\hsize}
\includegraphics[width=7cm]{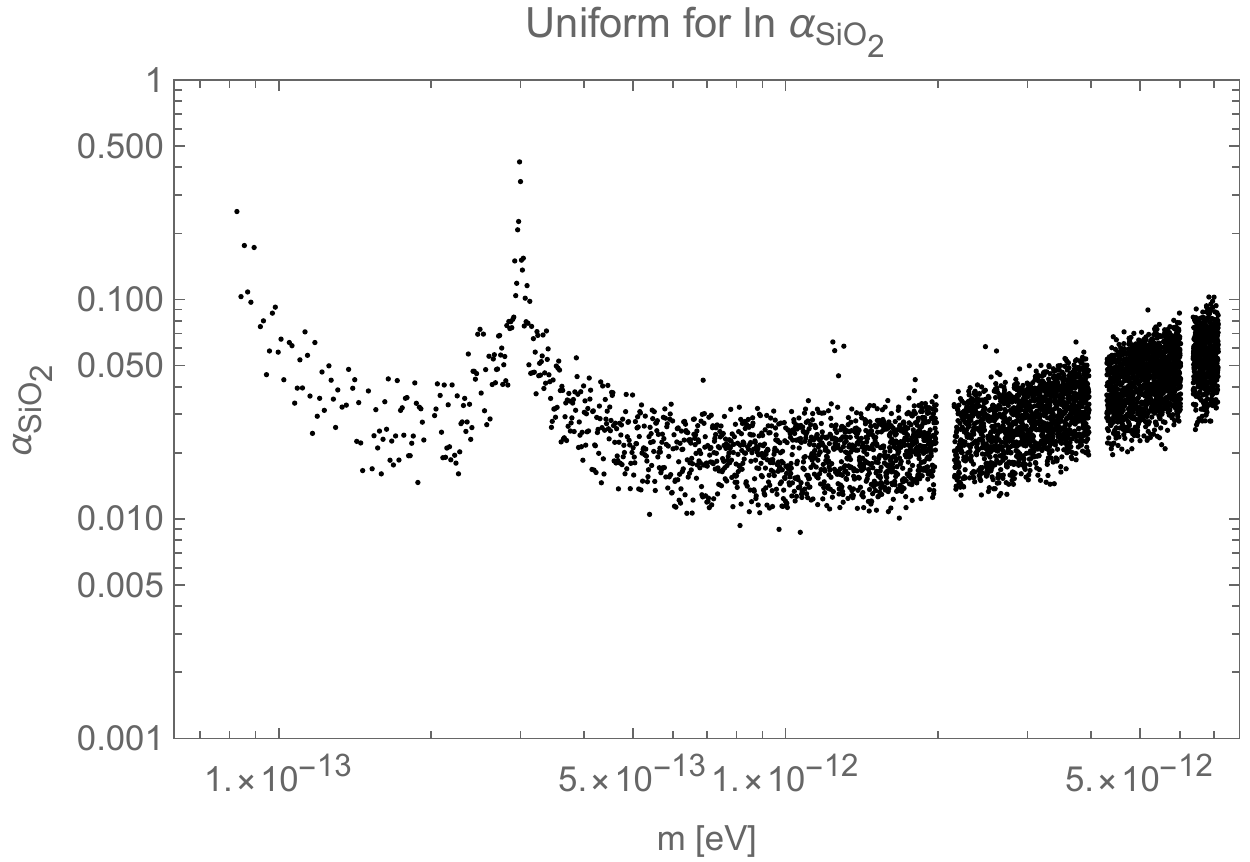}
\end{minipage}
\end{tabular}
\caption{Upper limit on $\alpha_{Si O_2}$ obtained by LIGO O3 data.
The left/right panel assumes uniform prior for $\alpha_{Si O_2}/ \ln \alpha_{Si O_2}$.}
\label{upper-limit}
\end{center}
\end{figure}

Fig.~\ref{upper-limit} is one of the main results.
Each black dot in both panels shows upper limit for corresponding value of $m$.
The left/(right) panel assumes uniform prior for $\alpha_{Si O_2}/ (\ln \alpha_{Si O_2})$.
Although the original data of the cross correlation statistics is given with 
frequency interval $\delta f \simeq 0.03~{\rm Hz}$,
we did not use data for all frequencies 
but picked up data only at frequency interval $10\delta f$ 
in order to reduce the file size of Fig.~\ref{upper-limit}.
Vertical white bands in both panels where there are no black dots are due to deficit of data.
Spike at $m\simeq 3\times 10^{-13}~{\rm eV}$ reflects the effect of the motion
of the Earth with respect to the rest frame of the Milky Way galaxy (see 
discussion regarding Fig.~\ref{fig.Q}). 
Finite spread of the black dots (i.e., that difference of values of $y$-axis between 
neighboring dots fluctuate) is caused by the fluctuations of the cross correlation data
that exist even among slightly different frequencies.
We find that there is no significant difference between the left panel and the right panel.
Thus, we use only the left panel for the following results.

\begin{figure}[htbp]
    \begin{tabular}{cc}
      \begin{minipage}[t]{0.45\hsize}
        \centering
        \includegraphics[keepaspectratio, scale=0.5]{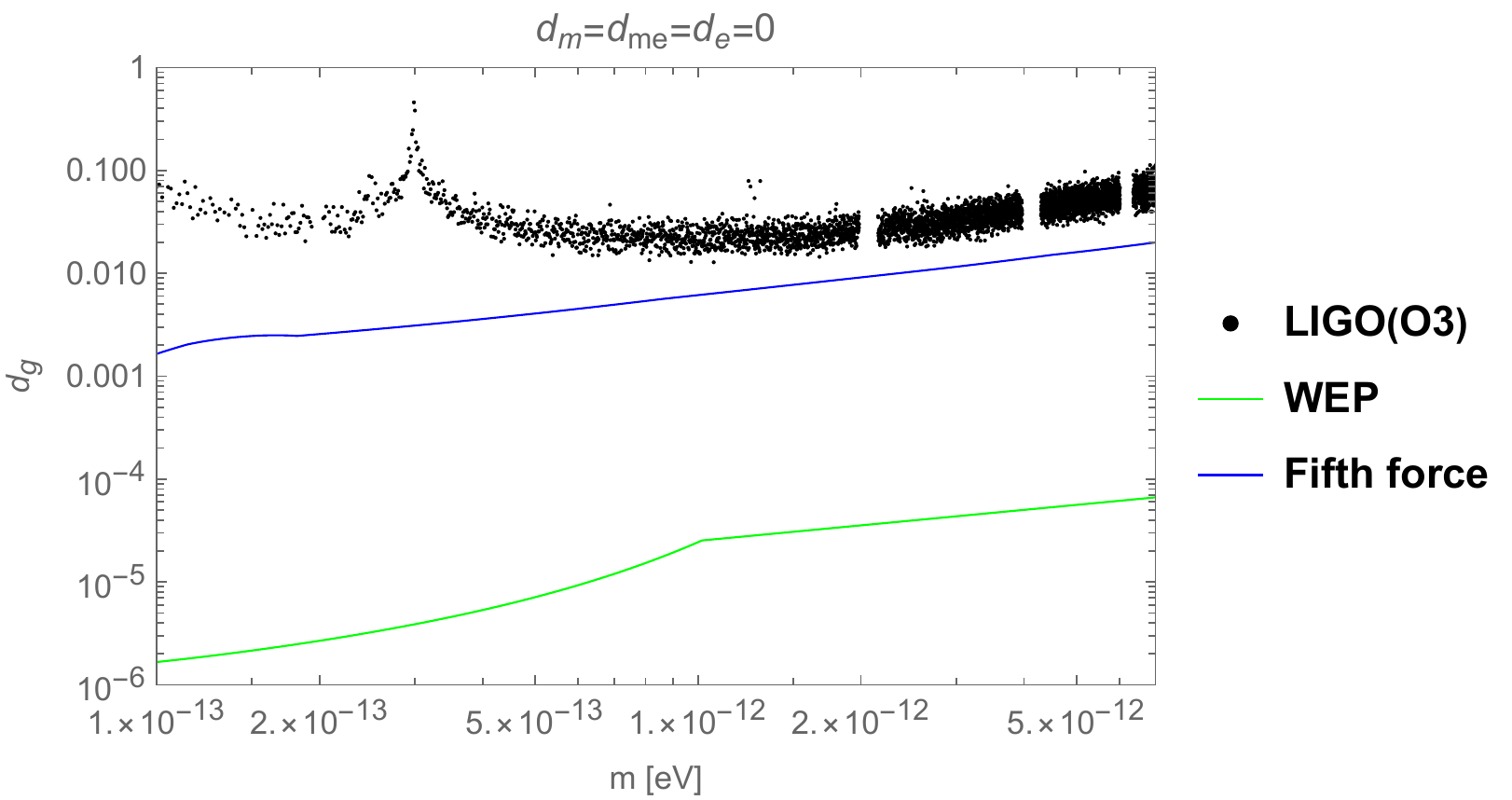}
      \end{minipage} &
      \begin{minipage}[t]{0.45\hsize}
        \centering
        \includegraphics[keepaspectratio, scale=0.5]{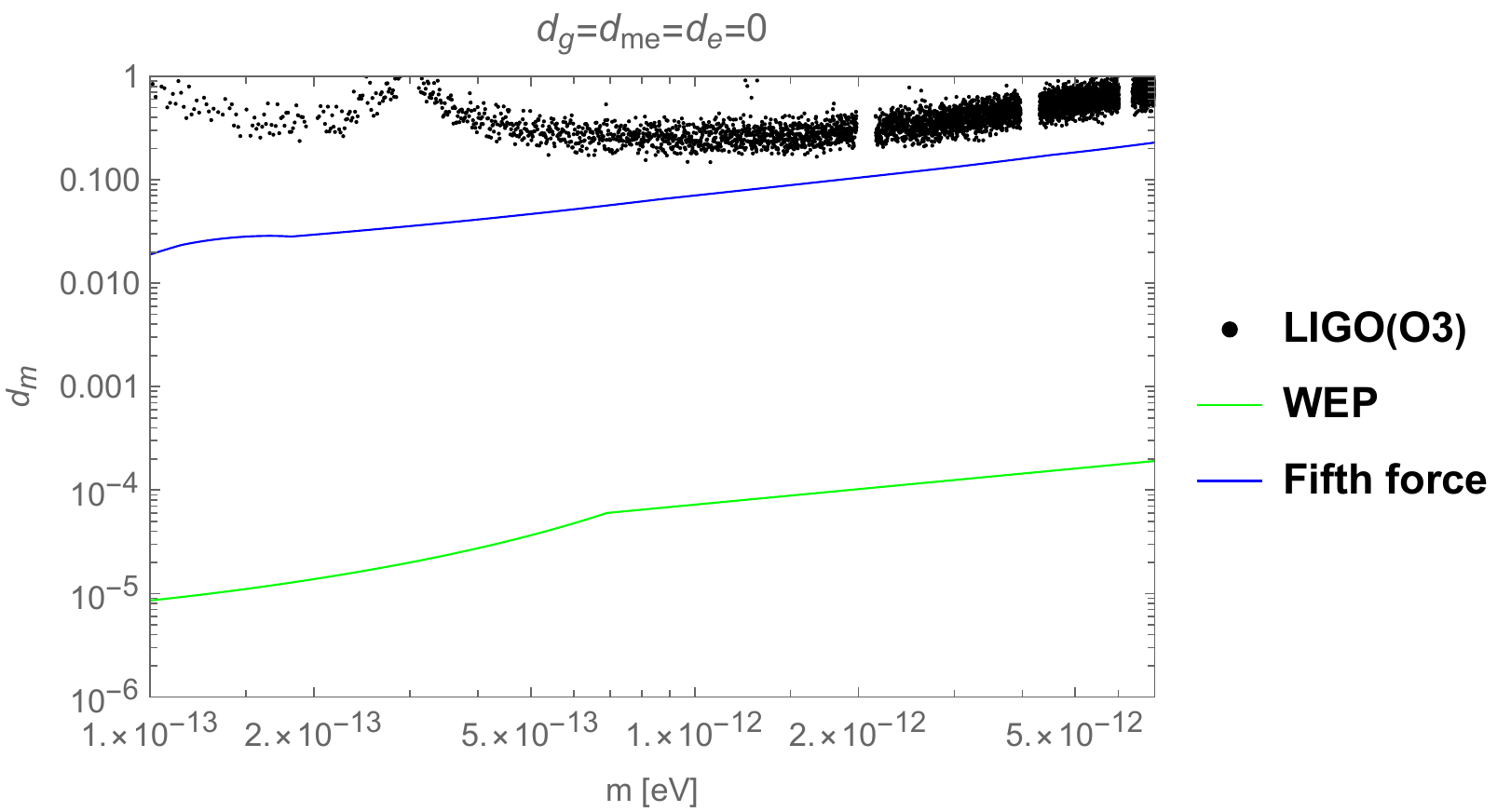}
      \end{minipage} \\
   
      \begin{minipage}[t]{0.45\hsize}
        \centering
        \includegraphics[keepaspectratio, scale=0.5]{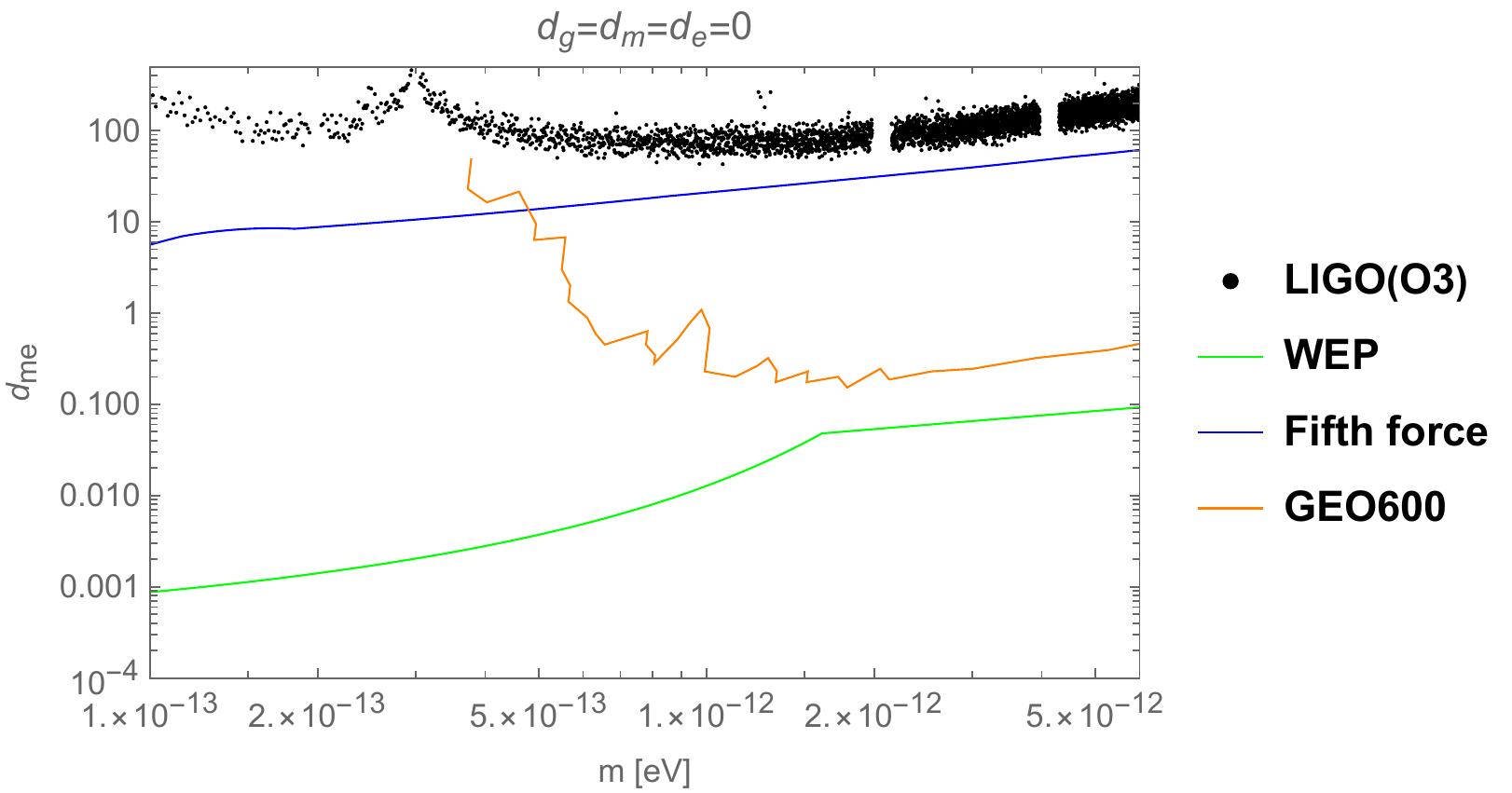}
      \end{minipage} &
      \begin{minipage}[t]{0.45\hsize}
        \centering
        \includegraphics[keepaspectratio, scale=0.5]{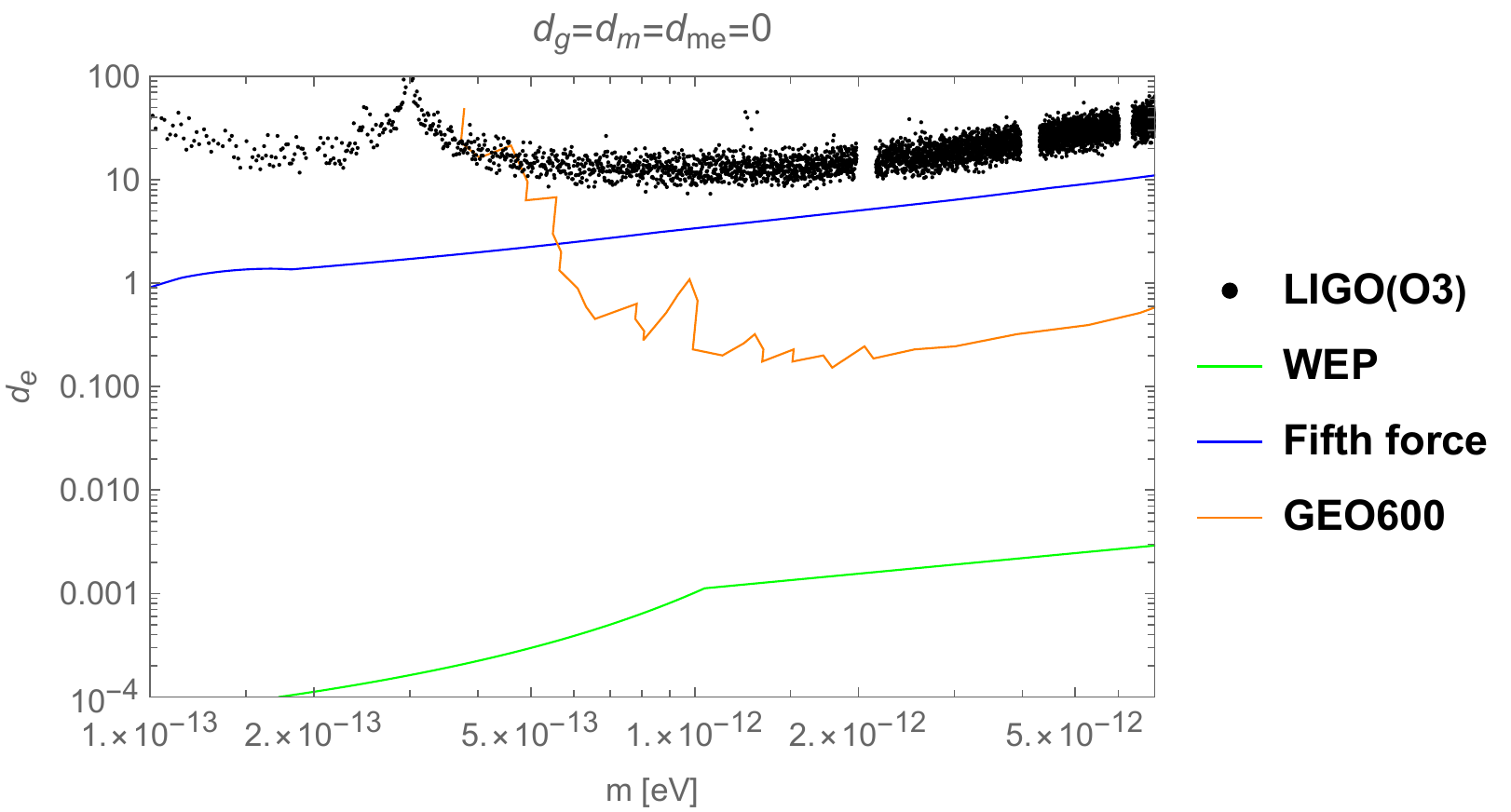}
      \end{minipage} \\
      \begin{minipage}[t]{0.45\hsize}
        \centering
        \includegraphics[keepaspectratio, scale=0.5]{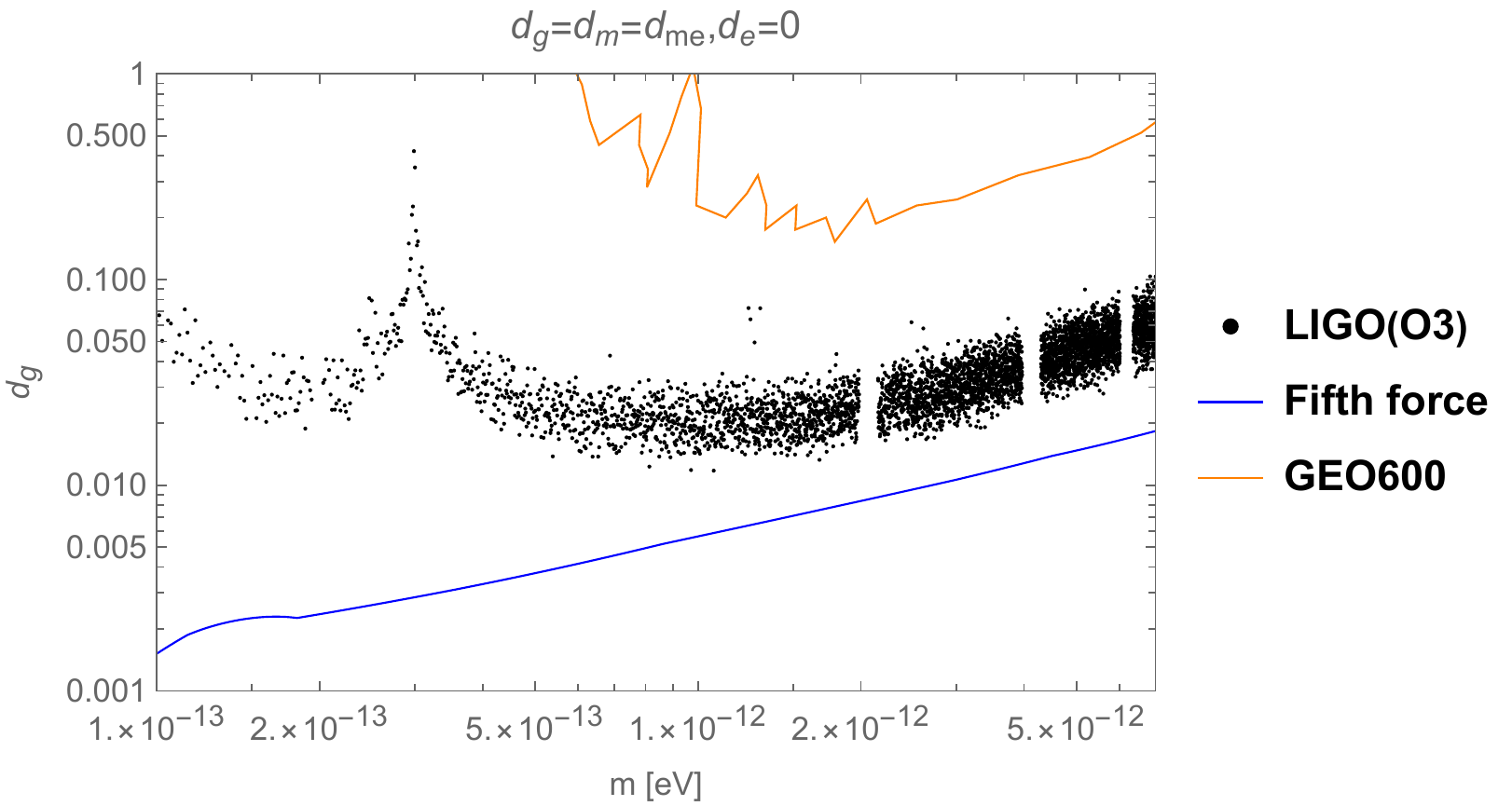}
      \end{minipage}
    \end{tabular}
    \caption{Upper limits on the coupling constants for five representative cases 
    which have been obtained by translating
    Fig.~\ref{upper-limit} using Eq.~(\ref{alpha-sio2}).
    Top left panel: upper limit on $d_g$ assuming $d_{\hat m}=d_{m_e}=d_e=0$.
    Top right panel: upper limit on $d_{\hat m}$ assuming $d_g=d_{m_e}=d_e=0$.
    Middle left panel: upper limit on $d_{m_e}$ assuming $d_g=d_{\hat m}=d_e=0$.
    Middle right panel: upper limit on $d_e$ assuming $d_g=d_{\hat m}=d_{m_e}=0$.
    Bottom panel: upper limit on $d_g$ assuming $d_g=d_{\hat m}=d_{m_e}, d_e=0$ which
    is the model of the gravitational non-minimal coupling discussed in the Appendix.}
\label{upper-d}
  \end{figure}

Fig.~\ref{upper-d} shows translation of Fig.~\ref{upper-limit} into the upper limits on the 
four coupling constants $(d_g,d_{\hat m}, d_{m_e}, d_e)$ for five representative cases 
by using Eq.~(\ref{alpha-sio2}).
In addition to our constraint by LIGO O3 data, we also show constraints from another gravitational-wave (GW)
experiment (GEO600) \cite{Vermeulen:2021epa}, which measures change of the size of the beam splitter
and is sensitive only to $d_{m_e}$ and $d_e$, as well as other non-GW experiments including 
test of violation of the weak equivalence principle (WEP) \cite{Schlamminger:2007ht, Berge:2017ovy, MICROSCOPE:2022doy}
and the fifth-force experiment \cite{Adelberger:2003zx}.
Experiments of WEP such as MICROSCOPE mission and the torsion balance are only sensitive to the difference of gravitational accelerations between different materials.
The effect of the violation of WEP is normally expressed by the Eotvos parameter $\eta$
defined by \cite{Berge:2017ovy}
\be
\eta = \frac{2(a_A-a_B)}{a_A+a_B}= \alpha_E (\alpha_A-\alpha_B) \Phi (mR_E)
\left( 1+mr \right) e^{-mr} \propto ( \alpha_A-\alpha_B ).
\ee
Here $a_A/(a_B)$ is the gravitational acceleration of material A/(B) measured at 
distance $r$ from the center of the Earth
and $\alpha_E$ is $\langle \alpha_A \rangle $ for materials constituting the Earth.
$R_E$ is the mean Earth radius and $\Phi (x) \equiv 3 (x\cosh x-\sinh x)/x^3$.
In our analysis, we assume that the Earth is made of silica (50\%) and magnesium oxide (50\%).
For the MICROSCOPE mission, A and B are beryllium (Be) and Platinum (Pt).
For the torsion balance experiment, A and B are beryllium (Be) and Titanium (Ti).
The fifth force experiments measure extra force operating between two materials 
in addition to the gravitational force.
For a pair of material A and the Earth, the experiment can probe a direct product 
$\alpha_A \alpha_E$.
We make an approximation that the material used in \cite{Adelberger:2003zx} is purely aluminium (Al).
The quantity $\alpha_{\rm Al} \alpha_E$ does not vanish for all the five cases in Fig.~\ref{upper-d}.
Thus, the upper limit set by the fifth-force experiment is present in all the panels
and is shown as a blue curve.

Let us now consider the upper limit for each case one by one.
The fist case in which only $d_g$ is assumed to be non-vanishing is shown in
the top left panel.
Since both $d_{m_e}$ and $d_e$ are zero, GEO600 does not give any meaningful constraint in this case.
Since the component-dependent part of Eq.~(\ref{alpha_A}) contains $d_g$,
WEP experiments can probe $d_g$.
The upper limit set by WEP is shown as a green curve.
We find that WEP provides the most stringent constraint by several orders of magnitude stronger
than the other constraints.
Even the LIGO at design sensitivity will not be able to reach the current constraint by WEP.
However, Cosmic Explorer, planned third GW detector, 
may be able to have better sensitivity than WEP \cite{Morisaki:2018htj}.
The second case in which only $d_{\hat m}$ is assumed to be non-vanishing is
shown in the top right panel.
This case is similar to the first case except that all the constraints are weaker than the ones
in the first case.
This is because $\alpha_E$ is less sensitive to $d_{\hat m}$
than $d_g$, namely $\alpha_E \simeq 0.9d_g/(0.08d_{\hat m})$ for the first/(second) case.
The third case in which only $d_{m_e}$ is assumed to be non-vanishing is shown
in the middle left panel.
In this case, GEO600 gives stronger constraint than LIGO for $m\gtrsim 3\times 10^{-13}~{\rm eV}$.
The WEP still provides the tightest upper limit on $d_{m_e}$.
The forth case in which only $d_e$ is assumed to be non-vanishing is shown
in the middle right panel.
This case is similar to the third case except that WEP is much stronger.
This can be understood from Eq.~(\ref{alpha_A}) that component-dependent term
proportional to $d_{m_e}$ is suppressed by a factor $(A-2Z)/A$ compared to
$d_e$.
The fifth case in which $d_g=d_{\hat m}=d_{m_e}$ and $d_e=0$ is assumed
is shown in the bottom panel.
In this case, WEP constraint is not present. 
The LIGO (O3) constraint is weaker than the fifth force experiment
although the difference is less than a factor of $\sim 5$ at large mass side.
The upgraded LIGO observation is expected to exceed the fifth force constraint \cite{Morisaki:2018htj}.

\section{Conclusion}
If dark matter is a light scalar field weakly interacting with elementary particles,
such a field induces oscillations of the physical constants, which results in
time-varying force acting on macroscopic objects.
We searched for the signal of the scalar field in the latest data of LIGO observations (O3 run)
and placed upper limits on the coupling strength between the mirrors in the laser
interferometers and the scalar field.
To this end, we first formulated the cross-correlation statistics 
that can be readily compared with publically available data.
It has been found that inclusion of the anisotropies of the velocity distribution of dark matter 
enhances the signal by a factor of $\sim 2$ except for the narrow mass range around
$\simeq 3\times 10^{-13}~{\rm eV}$ for which the correlation between the interferometer
at Livingston and the one at Hanford is blind to the signal.

From the non-detection of the signal, we also derived the upper limits on the coupling constants
between the elementary particles and the scalar field for five representative cases.
For all the cases where the weak equivalence principle is not satisfied,
tests of the violation of the weak equivalence principle provides the tightest upper limit
on the coupling constants.
Upper limits from the fifth-force experiment are always stronger than the ones
from LIGO, but the difference is less than a factor of $\sim 5$ at large-mass range,
which vividly demonstrates that the gravitational-wave experiments have been coming close
to the sensitivity of the fifth force experiment.
For the models where only the coupling to electrons or photos is non-vanishing,
upper limits from GEO600 are stronger than those from LIGO at large-mass range
while LIGO provides stronger limits for the case where the weak equivalence principle
is not violated.

Our study presents that gravitational-wave experiments are starting to bring
us meaningful information to constrain the nature of dark matter.
The formulation provided in this paper may be applied to the data of upcoming 
experiments as well and is expected to probe much wider parameter range of the model.

\section*{Acknowledgements}
We would like to thank Yuta Michimura for helpful comments on the manuscript.
This work is supported by the MEXT KAKENHI Grant Number 17H06359 (TS), JP21H05453 (TS),  
and the JSPS KAKENHI Grant Number JP19K03864 (TS).

\appendix

\section{Non-minimal coupling model}
We assume that the scalar field dark matter does not directly interact with standard model fields
and is only non-minimally coupled to gravity through the kinetic term mixing between the gravitons 
and the scalar field.
In the Jordan frame, the action is defined as
\be
\label{action-non-minimal}
S[{\tilde g}_{\mu \nu}, \Phi, \Psi]=\int d^4x~\sqrt{-{\tilde g}}
\left( \frac{1}{16\pi G}{\tilde R}-\frac{1}{2} {\tilde g}^{\mu \nu} \partial_\mu \Phi
\partial_\nu \Phi -{\kappa}^{-1} \zeta {\tilde R} \Phi-\frac{M^2}{2} \Phi^2 \right)
+S_m [ {\tilde g}_{\mu \nu}, \Psi].
\ee
Here ${\tilde g}_{\mu \nu}$ is the Jordan frame metric, 
scalar field, respectively.
The standard model fields are symbolically denoted by $\Psi$.
The mass term is parametrized by $M$ and 
dimensionless parameter $\zeta$ represents the strength of the kinetic term mixing.

In the standard manner \cite{Maeda:1988ab}, we can move to the Einstein frame where the gravitational action
takes the Einstein-Hilbert form by the following conformal transformation:
\be
{\tilde g}_{\mu \nu}=\frac{1}{1-4\zeta \kappa \Phi} g_{\mu \nu}.
\ee
In the Einstein frame, $\Phi$ is not a canonical field (i.e. coefficient of the kinetic term
is not $-\frac{1}{2}$) and the canonical field $\phi$ is obtained from $\Phi$ by
\be
\phi =\int_0^\Phi \frac{\sqrt{1+12\zeta^2-4\zeta \kappa x}}{1-4\zeta \kappa x}dx
=\sqrt{1+12\zeta^2}\Phi+{\cal O}(\Phi^2).
\ee
Then, at the leading order in $\phi$, the action in the Einstein frame becomes
\be
S[g_{\mu \nu}, \phi, \Psi]=\int d^4x~\sqrt{-g}
\left( \frac{1}{16\pi G}R-\frac{1}{2} g^{\mu \nu} \partial_\mu \phi
\partial_\nu \phi -\frac{m^2}{2} \phi^2 \right)
+S_m [ \Omega^2 (\phi) g_{\mu \nu}, \Psi],
\ee
where the mass $m$ of $\phi$ and the conformal function $\Omega (\phi)$ are given by
\be
m=\frac{M}{\sqrt{1+12\zeta^2}},~~~~~~\Omega (\phi) =1+\frac{2\zeta}{\sqrt{1+12\zeta^2}} \kappa \phi 
+{\cal O}(\phi^2).
\ee
The presence of the conformal function leads to the dependence of the energy scale $E$
inherent in $S_m$ on the value of $\phi$ as $E(\phi)= \Omega (\phi) E$.
Consequently, the low-energy physical constants listed in Eqs.~(\ref{alpha-phi})-(\ref{mud-phi})
depends on $\varphi$ as
\begin{align}
&\alpha (\varphi )=\alpha, \\
&\Lambda_3 (\varphi )=\left( 1+\frac{2\zeta}{\sqrt{1+12\zeta^2}} \varphi \right) \Lambda_3, \\
&m_e (\varphi )=\left( 1+\frac{2\zeta}{\sqrt{1+12\zeta^2}} \varphi \right) m_e, \\
&m_{u,d}(\varphi )=\left( 1+\frac{2\zeta}{\sqrt{1+12\zeta^2}} \varphi \right) m_{u,d}(\Lambda_3).
\end{align}
Thus, the model defined by Eq.~(\ref{action-non-minimal}) corresponds to a set of parameter values
\be
d_g=d_{m_e}=d_{m_{u,d}}=\frac{2\zeta}{\sqrt{1+12\zeta^2}},
~~~~~~~d_e=0.
\ee

\bibliography{ref}
\end{document}